\documentclass{article}

\usepackage{arxiv}

\usepackage[utf8]{inputenc} 
\usepackage[T1]{fontenc}    
\usepackage{hyperref}       
\usepackage{url}            
\usepackage{booktabs}       
\usepackage{amsfonts}       
\usepackage{nicefrac}       
\usepackage{microtype}      
\usepackage{graphicx}
\usepackage{doi}
\usepackage{blindtext}
\usepackage{enumitem}
\usepackage{epsfig}
\usepackage{graphicx}
\usepackage{amsmath}
\usepackage{amssymb}
\usepackage{multicol}
\usepackage{float}
\usepackage{mcite}
\usepackage[euler]{textgreek}
\usepackage{subcaption}
\usepackage{physics}
\usepackage{siunitx}
\usepackage{cite}

\DeclareSIUnit\molar{\mole\per\cubic\deci\metre}
\DeclareSIUnit\Molar{\textsc{m}}

\setitemize[0]{leftmargin=*}
\setenumerate[0]{leftmargin=*}

\renewcommand{\headeright}{}
\renewcommand{\undertitle}{}
\renewcommand{\shorttitle}{A single molecule MCB junction in the partially wet phase.}

\hypersetup{
pdftitle={},
pdfsubject={},
pdfauthor={},
pdfkeywords={},
}

\begin{document}
\title{detailing the natural creation process of a single molecule MCB junction in the partially wet phase}

\author{C.J. Muller\thanks{Correspondence: PO Box 60, 6580 AB Malden, The Netherlands}
\\ \\
This research has, in its entirety, been privately conducted and funded by the author.
}
\maketitle

\begin{abstract}
A novel way to arrive at a single molecule junction in the partially wet phase is described. The fabrication of this device makes use of a natural drying process. Details are provided concerning the fabrication of the device and the related material choices. In addition a process description is provided with a step by step approach for creating the partially wet phase molecular MCB junction.
\end{abstract}

\begin{multicols}{2}

\section{Introduction}
The wetting of surfaces is an interesting and current field where chemistry, physics and engineering come together \cite{bonn2009source1}. Chemistry ensures interaction at the molecular scale. Van der Waal forces extend over tens of molecules providing a larger interaction scale. The engineering comes in by creating for example repetitive structures thus providing additional length scales as a means to enhance our understanding of specific wetting situations.

The partially wet phase layer of tetrahydrofuran fluid is a requirement for observing a quantum effect related to continuous Q-charging at ambient conditions \mcite{muller2021source2,*muller2021source2b}.  This is only adding to the interest in the field of wetting, in this case at the microscopic level. The fundamental question is how a microscopic layer of fluid will wet a single molecule anchored between two electrodes and how this layer behaves over time. This system is in analogy to the terminology in the field of wetting \cite{bonn2009source1} called a partially wet phase molecular MCB junction.

Due to a large number of, only in part controlled, degrees of freedom associated with the partially wet phase molecular MCB junction a number of critical material decisions need to be made. Each of these decisions can lead to failure when not carefully investigated. The experimenter will be confronted with many unknowns and uncertainties, to mention a few:

\begin{itemize}
    \item How to end up with one molecule bridging the electrodes without any positional dependence of the IV curve on the gap within a fluid.
    \item Is the solvent used in the cell pure enough and not contaminating the electrodes.
    \item Are the molecules used in the solution modified at ambient conditions.
    \item Is the seal-material or any other material inside the cell dissolving in the solution and perhaps contaminating the electrode material.
    \item Is any of the material inside the cell which is in contact with the solution, incompatible with or modifying the dissolved molecules. 
\end{itemize}
To find a suitable process-window and way of work in the above mentioned parameter space can be tedious and laborious. It is the aim of this paper to enable those who are trying to replicate and build on the earlier reported results in this field \cite{muller2021source2}. The limitation will be that the MCB assembly as well as the solution and molecule choice is restricted to what the author has been using: a solution of anhydrous tetrahydrofuran (THF) with dissolved benzene di-thiol (BDT) molecules or a pure THF fluid in combination with a notched filament MCB assembly. In addition H\textsubscript{2}O can be used in combination with the aforementioned possibilities.

Macroscopic wetting layers may affect micro fabricated bridges in a different manner as compared to their notched filament MCB counterparts, simply because the gaps where fluid can be trapped are smaller. Trapped fluid may influence/exert forces on free hanging bridges. In order to be able to mimic the experimental situation in reference \cite{muller2021source2}, the entire notched filament device fabrication is described in detail. For operating the device the use of a pivoting MCB setup \cite{muller2021source2} can be controlled in case a \SI{4.5}{\milli\meter} inner diameter cell ($\phi_\text{in}^{4.5}$) is used. This cell will release the fluid to the cell-opening under pivoting angles where it can be absorbed.  Also a cell with a smaller diameter can be used in a static MCB setup in this case the draining will take place via a paper tip. Many devices with a cell inner diameter of \SI{2}{\milli\meter} ($\phi_\text{in}^2$) have successfully been operated.

Since the number of junction types as well as partially wet phase liquids will be limitless the following naming convention is proposed. The barrier molecule forms the base and the lining partially wet phase forms the exponent. For example BDT\textsuperscript{THF} for a BDT barrier junction in the THF partially wet phase and BDT\textsuperscript{H2O} for a BDT barrier junction in the H\textsubscript{2}O partially wet phase.

\section{Device fabrication}
The basis for the device is a $1\times5\times24$~\si{\milli\meter} phosphor bronze bending beam. The bending beams have been laser cut, are wet sanded and aceton ultrasonic cleaned prior to a furnace step of \SI{130}{\celsius} for 3 hours. This will discolor the bending beams from a shiny metallic color to an oily film like color. This step enhances the adhesion of the Stycast glue layer used to surface the bending beam with Kapton. The Kapton foil of \SI{200}{\micro\meter} thick is glued on the bending beam, using Stycast 2850 with \SI{3.5}{\percent} CAT9 catalyst. This combination is favored over other Stycast catalyst combinations, as after ambient curing it results in the most chemical resistant epoxy (amongst the Stycast family). In addition it caters for a relative high temperature furnace cure, which further enhances chemical resistance. 

For the electrode material a filament of hard temper Au (\SI{25}{\micro\meter}) or a filament of annealed soft temper Au (\SI{75}{\micro\meter}) has been used.  The filament is notched by a scalpel on a lathed stainless steel surface finish (e.g. bottom of a pan), making use of the minute grooves which cater for not fully cutting through the filament. Subsequently half of the filament is fixated on this surface by covering it with a piece of metal; the notch is "fatigued" by moving the other half up and down until it almost breaks. The weakened notched filament is positioned on the Kapton and fixated at the outer ends using Stycast. For the fixation of the center notched section again the same Stycast catalyst combination is used. The creation of the central Stycast contacts is performed with half cured Stycast, 2 hours and 45 minutes after mixing of the compound at room-temperature. About \numrange{20}{30} minutes are available for creating the central Stycast contacts on a hot plate at approximately \SI{50}{\celsius}, under a microscope. The Stycast has hardened considerable, however can still be picked up. Holding it in close proximity to the hot plate will make it malleable again and allows it to be used for the center epoxy contacts. The central unglued section can be fine-tuned by the use of heat to let the Stycast contacts flow wider or by taking the assembly off the hot plate to stop the slowly flowing Stycast contacts. Alternatively the epoxy contacts can be pushed closer together while the assembly rests on the hot plate. After the MCB assembly, including the filament fixture, is cured at ambient conditions it will undergo a 3 hour \SI{130}{\celsius} furnace cure in order to enhance chemical resistance of the Stycast contacts. Ramp up and down should be slow, 15 minutes, in order to prevent breaking filaments.

After the creation of the bending beam construction, the liquid cell needs to be assembled. A glass cell is used, which is simply a piece of glass tube cut to the right size, approximately \SI{25}{\milli\meter} has been used. One opening of the cell should encircle the notched filament. The liquid rubber will after curing ensure a leak tight electrical feedthrough. Furthermore the elastomeric seal provides a leak tight connection between the Kapton layer and the glass cell; it also caters for flexing of the bending beam. Ideally the liquid rubber seal does not extend higher than \SIrange{1.5}{2}{\milli\meter} above the bending beam for the $\phi_\text{in}^2$ cell. This is to ensure the possibility during the draining process to look from the side of the cell to the inside. The low seal height can be achieved as follows:

\begin{itemize}
    \item Fixate the cell to the bending beam using an extremely thin ribbon of liquid rubber at the outer diameter bottom side. Liquid rubber preferably does not enter the inside of the cell. This is only for fixation, this step will in general not deliver a leak tight seal.
    \item Let the assembly cure for approximately an hour.
    \item Apply some tape circling the cell, exposing the bottom 1.5-2~\si{\milli\meter}.
    \item Apply a big blob of liquid rubber circling the cell within 5 seconds in one go making sure all exposed surfaces are well contacted.
    \item Tearing off the tape leaves a low height seal.
    \item The assembly needs to cure again, this time following the specifications related to the liquid rubber.
\end{itemize}

\begin{figure}[H]
    \centering
    \includegraphics[width=0.48\textwidth]{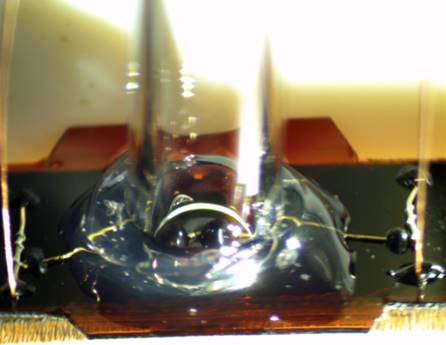}
    \caption{A photo of the bending beam assembly. Visible are the notched section with Stycast contacts, the elastomeric seal, the cell and the electrical feedthroughs. The notched gold wire is connected to two copper wires which are connected to the wiring of the MCB setup.}
    \label{fig:1}
\end{figure}

After application of the glass cell and curing of the liquid rubber at ambient conditions, the entire assembly gets another furnace cure for 1 hour at \SI{110}{\celsius}, in this case to enhance the chemical resistance of the seal. For the seal Bison Rubber Repair/liquid rubber is used, which is a transparent, solvent free rubber paste from a tube. It cures under the influence of air humidity and is chemically resistant. Fig.~\ref{fig:1} shows a photo of the $\phi_\text{in}^2$ cell. The THF/BDT solution in the cell is in contact with the following materials:

\begin{itemize}
    \item Gold electrode and filament material
    \item Stycast, the center notched filament epoxy contacts
    \item Kapton, covering and insulating the bending beam
    \item Glass, the capillary used as a cell
    \item Liquid rubber, in use as an elastomeric seal
\end{itemize}

It can never be excluded that some traces of the used materials get dissolved in the THF fluid, however reproducible measurement differences with respect to THF\textsuperscript{THF} and BDT\textsuperscript{THF} junctions have been obtained with this material combination \cite{muller2021source2}.

\section{The formation process of a molecular junction}
Draining and the subsequent drying process is critical in obtaining the partially wet phase junction, be it the BDT barrier or the THF barrier junction. The formation of the molecular junction is initiated by the drying process. In addition the right timing is required for starting the measurements in the partially wet phase. In reference \cite{muller2021source2} is has been phrased as follows:

"A law of nature seems to hold for two electrodes spaced at nanometers distance during the drying process. Be it soft temper Au or hard temper Au used as the MCB electrode material, be it broken in air and after that immersed in fluid or broken directly in a THF environment, the shown electrode behavior upon the drying process is the same. Once the excess fluid has been drained, during the drying process the conductivity continuously increases. This is attributed to a continuously decreasing distance between the two closest points on the opposing electrodes, due to the lasting evaporation of the fluid on and between the electrodes. Apparently there is enough elasticity in the nanometers separated electrodes such that the exact initial separation does not matter. The continuously decreasing amount of fluid between the electrodes pulls the electrodes ever closer together. This process endures until the fluid is getting depleted, and a molecule is squeezed between the electrodes, blocking any further movement. The adhesion forces of the molecules (either BDT or THF) to the gold electrode surface, together with the partially wet phase layer which is lining the entire electrode-molecule-electrode system, are such that a metal-molecule-metal junction is favored over a metallic junction."

Although in the completely wet phase it is not possible to derive any positional dependence of the electrodes by electrical measurements during the drying process the situation is better. Since the conductance increases continuously during the drying process it is possible to monitor the conductivity over time. Also IV curves can be continuously recorded. It is however a possibility that a measurement influences the creation of a partially wet phase layer. Ideally in the end a partially wet phase layer of a uniform thickness is lining the molecular device. Current or electrical fields may exert forces on fluid molecules which may preferably occupy available space near the junction and get trapped there because of the geometry. Most of the data in reference \cite{muller2021source2} has been obtained by performing single point current measurements during the drying process at \SI{-5}{\volt} every few minutes or so and only start continuous IV scan measurements once the partially wet phase has formed. The exact trigger when to start the continuous IV measurements relies on both the current value at \SI{-5}{\volt} and a visible check of the cell. 

\section{Draining and relevant current values}
Just after breaking in either pure THF fluid or in a \SI{1}{\milli\Molar} BDT/THF solution any current value at \SI{-5}{\volt} between \SI{-5}{\nano\ampere} and \SI{-50}{\nano\ampere} is the norm. Occasionally values of \SI{-1}{\nano\ampere} or smaller are measured in the completely wet phase. Although there is no evidence it is suspected to be due to H\textsubscript{2}O molecules adhering to the surface of the electrodes. This cannot be undone. Continuing with the drying process will lead to a reduced end value of the current and degraded results as compared to a junction that showed normal start current values.

Draining of the $\phi_\text{in}^{4.5}$ cell is simply performed by pivoting the setup \cite{muller2021source2} absorbing the fluid at the cell opening and reverting back to the vertical cell position. For a $\phi_\text{in}^{2}$ cell the draining has to be performed by a paper tip and is more laborious. The last 3-4~\si{mm} of the tip needs to be sharply bent in order to eliminate the possibility of touching the junction in the center. In this case the end of the tip will touch the glass wall on the inside. During draining of the last \SI{2}{\milli\meter} fluid the inside of the cell needs to be monitored, as soon as the paper tip attracts fluid this will be visible and the tip needs to be retracted immediately. This is the reason for having a 1.5-2~\si{mm} seal-height it allows monitoring of the cell bottom from the outside during draining. Every time the tip has been retracted a visual inspection by a magnifying glass of the cell bottom is required in order to understand the exact phase of the draining process the cell is in. Once the desired phase in the drain process has been reached (see below) the cell needs to remain untouched. 

After draining the cell starts to gradually dry. It is important to note that the cell and the junction are coupled systems with respect to wetting. As long as the cell is wet, so is the junction. The cell acts as a buffer and the junction can only follow. This also implies that if it is possible to observe the last bit of fluid in the cell that this provides a good trigger for the start of the continuous IV curve measurements. This trigger becomes even more secure if it is possible to also tie it to a certain current value. The norm of the trigger value is \SI{-100}{\nano\ampere} at the \SI{-5}{\volt} value. At this value the current value in the IV curve measures 300-500~\si{\nano\ampere} at \SI{10}{\volt}. It is expected to represent the completion of the formation process of a single molecular junction. 

\section{Optical contour graphs as a means of determining the stage in the drain and drying process}
\begin{figure}[H]
    \centering
    \includegraphics[width=0.48\textwidth]{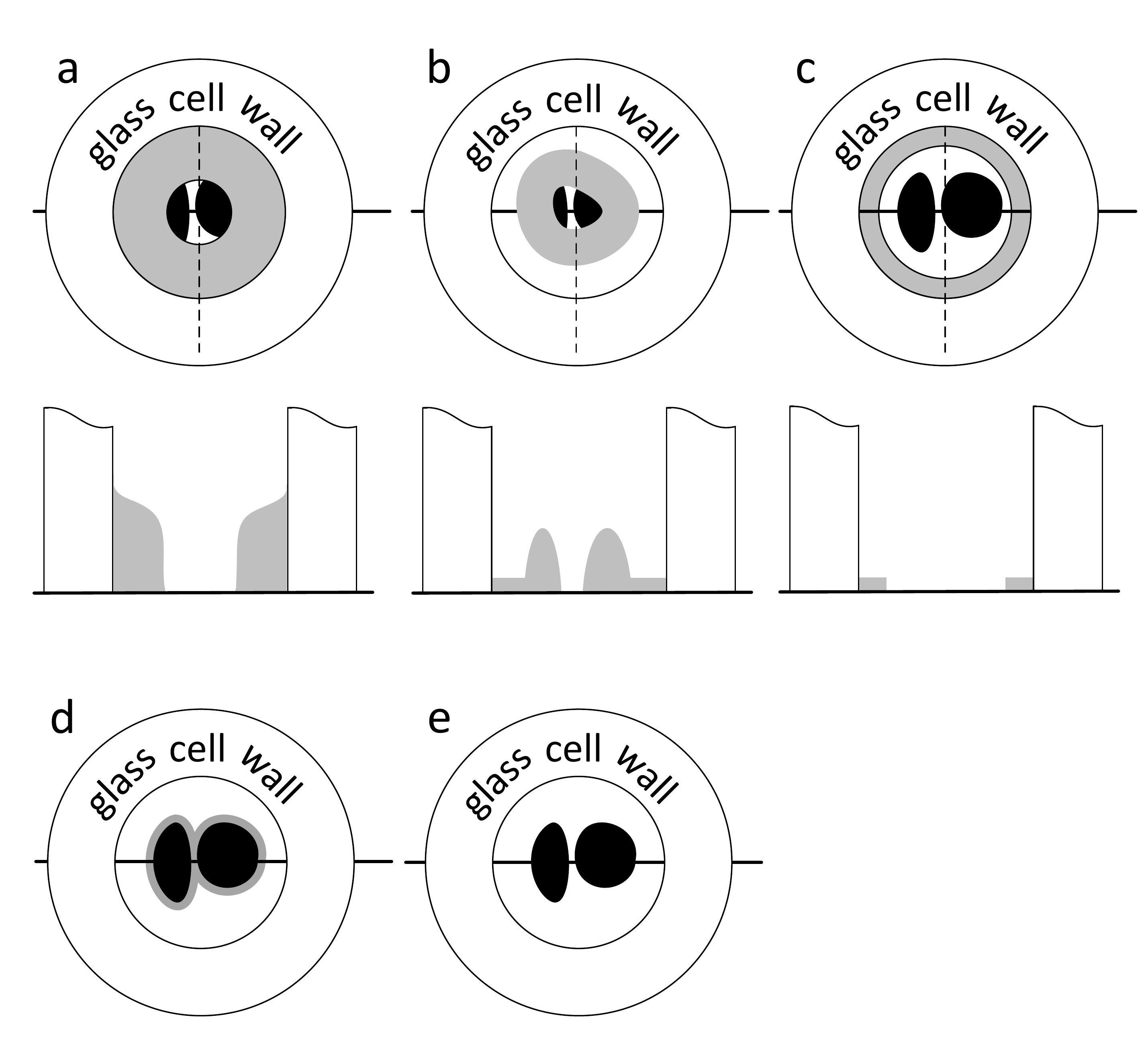}
    \caption{Optical contour graphs detailing the wetting geometry inside a \SI{2}{\milli\meter} inner diameter cell as a function of time. The THF contour in picture a may still last for hours. Paper tip draining should progress towards the contour in picture c from this stage it will take about 30 minutes to the completely dry phase. Picture d shows a THF liner around the Stycast contacts, once this liner is evaporated the partially wet phase is entered. Picture e provides the completely dry phase.}
    \label{fig:2}
\end{figure}

How to know what to look for during the draining process for the $\phi_\text{in}^{2}$ cell is indicated in Fig.~\ref{fig:2}. The draining via a paper tip should progress the experimenter from a situation depicted in Fig.~\ref{fig:2}a to Fig.~\ref{fig:2}c. The cross section of the fluid profiles is indicated below the top down contour graphs.

Initially when the bulk of the fluid is removed a situation indicated in Fig.~\ref{fig:2}a is encountered. It may still take hours to evaporate this amount, depending on ambient conditions. Carefully continuing draining leads at some point to a donut shaped THF ring pinned to the Stycast contacts, Fig.~\ref{fig:2}b. This ring is no longer touching the cell-wall. Continuing even further with the drain process will lead to an “outer ring” touching the cell-wall, Fig.~\ref{fig:2}c. This is the situation to stop draining and let the drying process progress to the partially wet phase in half an hour or so.

For both the $\phi_\text{in}^{2}$ cell and the $\phi_\text{in}^{4.5}$ cell the stage in the drying process just prior to the partially wet phase is as indicated in Fig.~\ref{fig:2}d. A liner of moisture encircles the Stycast contacts. This liner reduces in time until the cell is completely dry as in Fig.~\ref{fig:2}e. In practice the trigger to start continuous IV measurements is when the liner is almost invisible at the junction location. It is possible that fluid is still trapped at the edge where the inner cell wall contacts the Kapton. This will normally show as glare, which slowly degrades and may lengthen the life of the liner of moisture. In practice either the current value of \SI{-100}{\nano\ampere} at \SI{-5}{\volt} or the visual disappearance of the liner of moisture are used as a trigger to start the continuous IV scans. Using them both provides enhanced control.

\section{The BDT\textsuperscript{H2O} device as an enduring molecular junction at ambient conditions}
Enduring molecular junctions at ambient conditions are inherently lined with an H\textsubscript{2}O partially wet phase layer \cite{bonn2009source1}. In ref \cite{muller2021source2} it was reported that these types of junctions can be created starting off with the BDT/THF solution catering for a BDT\textsuperscript{THF} junction. The procedure to arrive at these interesting structures is detailed below. For all BDT\textsuperscript{H2O} junctions the $\phi_\text{in}^{2}$ cell has been used.

\begin{figure}[H]
    \centering
    \includegraphics[width=0.48\textwidth]{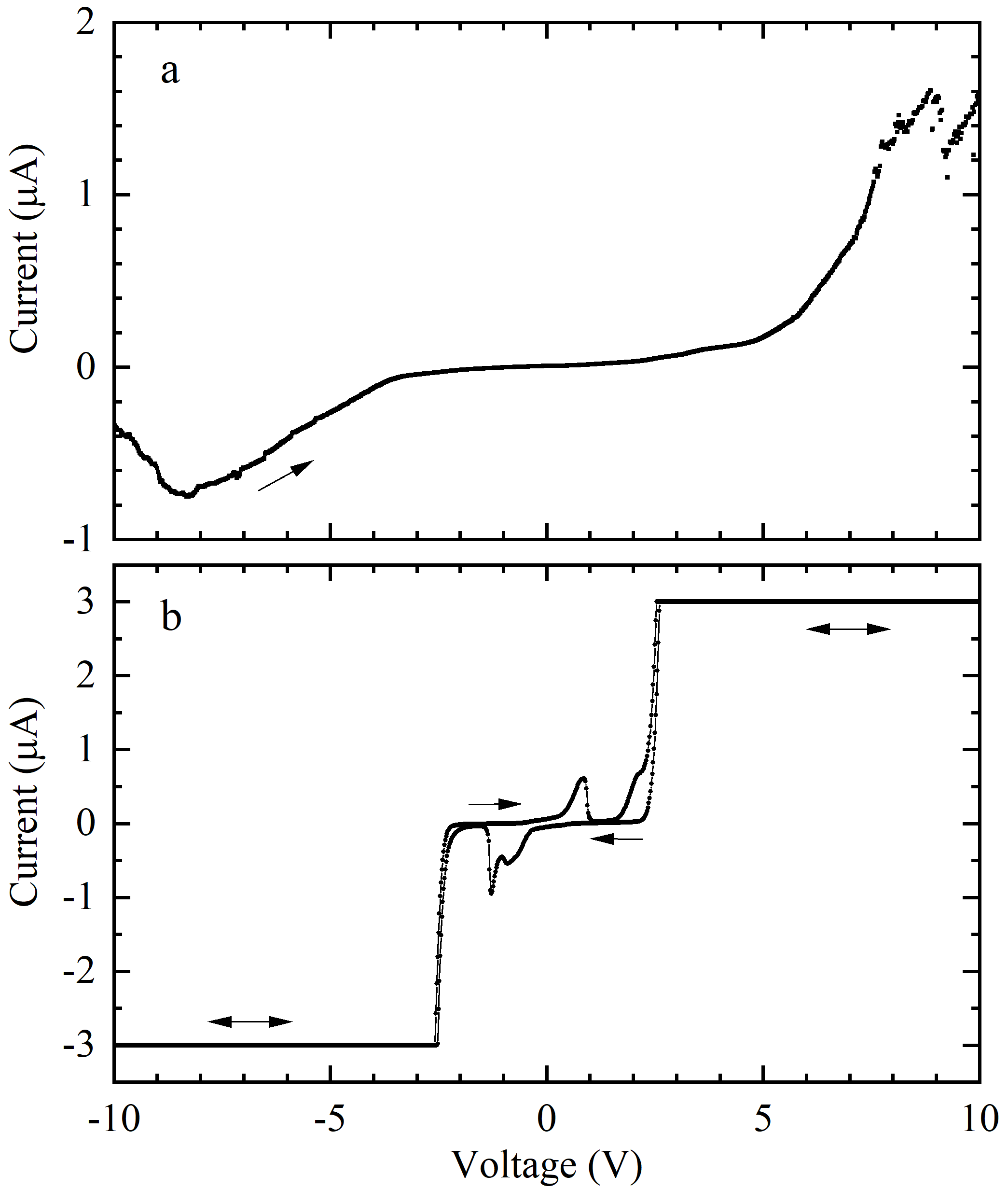}
    \caption{Progression to an enduring BDT\textsuperscript{H2O} junction by switching to \SI{95}{\percent} humidity towards the end of the BDT\textsuperscript{THF} formation process. The first resulting asymmetric IV curve is shown in panel a. Under successive IV trace recording after approximately half an hour the enduring IV curve as shown in panel b has been obtained.}
    \label{fig:3}
\end{figure}

The start to arrive at a BDT\textsuperscript{H2O} junction is by using the BDT\textsuperscript{THF} junction formation process. Towards the end of the formation process the humidity is enhanced to about \SIrange{90}{95}{\percent} by a humidifier in an enclosed space also containing the MCB setup. This is done in the \SIrange{-80}{-100}{\nano\ampere} range at \SI{-5}{\volt} in the BDT\textsuperscript{THF} formation process. Subsequently continuous IV curve scanning needs to start from \SIrange{-10}{10}{\volt} and reverse. Fig.~\ref{fig:3}a shows the initial IV curve of this process just after the humidity was raised. The current start level is about \SI{-400}{\nano\ampere} at \SI{-10}{\volt} and reaches values beyond \SI{1.5}{\micro\ampere} at \SI{10}{\volt} in a \SI{45}{\second} scan. From this stage onward the junction still modifies over the next half hour under continuous IV scanning (\SI{20}{\volt}/\SI{45}{\second}) to arrive at the final enduring IV curve which is shown in Fig.~\ref{fig:3}b.

The exact point in time when to switch the humidifier on is not critical. Also a BDT\textsuperscript{H2O} junction has been created by increasing the humidity to \SI{95}{\percent} after the cell dried and the conductivity was reduced to \SI{30}{\nano\ampere} at \SI{10}{\volt}. The subsequent scanning with a rate of \SI{20}{\volt}/\SI{5}{\second} lasted for approximately half an hour. At that point in time the enduring IV curve similar as in Fig.~\ref{fig:3}b was reached.

\begin{figure}[H]
    \centering
    \includegraphics[width=0.48\textwidth]{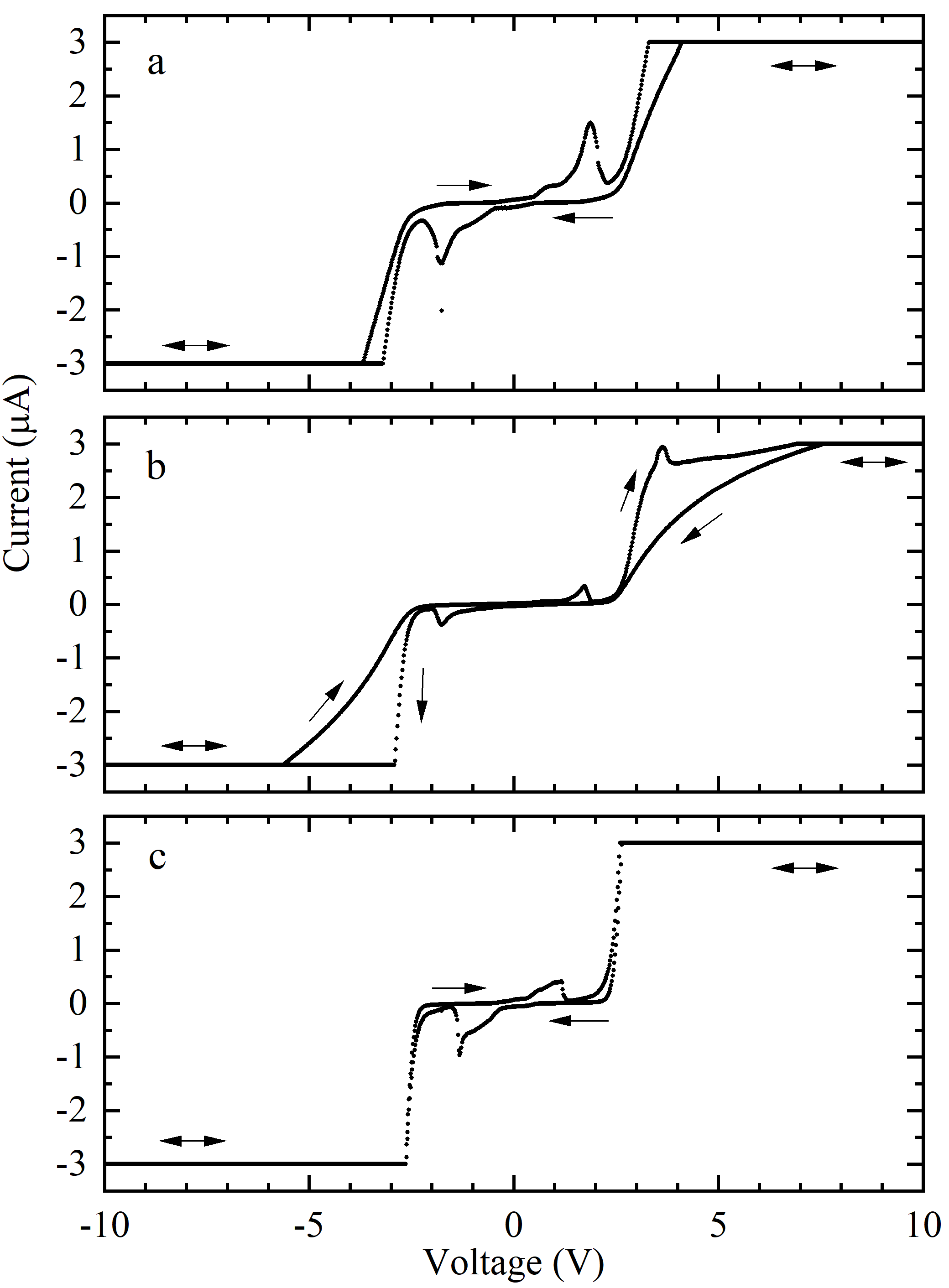}
    \caption{Progress of the formation of an enduring BDT\textsuperscript{H2O} junction from panel a to c. Panel a shows a junction which almost attains the enduring form. It has been achieved by continuous IV scanning. After stopping the continuous IV scans panel b showed the degradation of the junction as the outer legs are drifting apart. After a couple of minutes of stressing scans with a scan speed of \SI{4}{\volt\per\second} the final enduring situation has been reached in panel c.}
    \label{fig:4}
\end{figure}

The scanning is required for settling of the H\textsubscript{2}O partially wet phase molecules in the direct vicinity of the junction. The high electric fields exert changing forces on the H\textsubscript{2}O dipoles building the partially wet phase layer at the junction. This is shown in Fig.~\ref{fig:4} which displays successive stages of the junction in panel a to c. Panel a shows an IV curve which is not at the final enduring state. Panel b shows that the outer legs of the IV curve have moved apart as compared to panel a. After switching to a scan-rate of \SI{4}{\volt\per\second} between panel b and c for several minutes the two legs gradually pulled together to create the enduring IV curve as shown in panel c.

This is evidence that the IV measurements have an impact on the formation of a partially wet phase. The sub-gap structure was shown to be scan-speed dependent and is expected to result from H\textsubscript{2}O dipoles in the partially wet phase layer \cite{muller2021source2}. Scan speed values for the displayed curves in panel a, b and c  are \SI{4}{\volt\per\second}, \SI{0.44}{\volt\per\second} and \SI{0.44}{\volt\per\second} respectively over the entire \SIrange{-10}{10}{\volt} range.

The remarkable aspect about this junction is the stability over hours, after the initial high humidity is gradually reduced to ambient values. Multiple attempts on one assembly have been executed as long as the conduction remained above \SI{-5}{\nano\ampere} at \SI{-5}{\volt}.

Four out of eight assemblies led to the same IV characteristic shown in Fig.~\ref{fig:3}b and Fig.~\ref{fig:4}c after the BDT\textsuperscript{H2O} junction was completed. In analogy to the reasoning in the early 90’s why single atom point contacts exist, the consistent measured value close to $2e^2/h$ of the last conductance plateau has been a convincing argument see for example \cite{zhou1995microfabrication}. A similar reasoning holds for molecular junctions. Using the same type of molecules ending up in the same reproducible and enduring IV curves is a very strong indication that the same single molecular junction is created over and over again. 

\begin{figure}[H]
    \centering
    \includegraphics[width=0.48\textwidth]{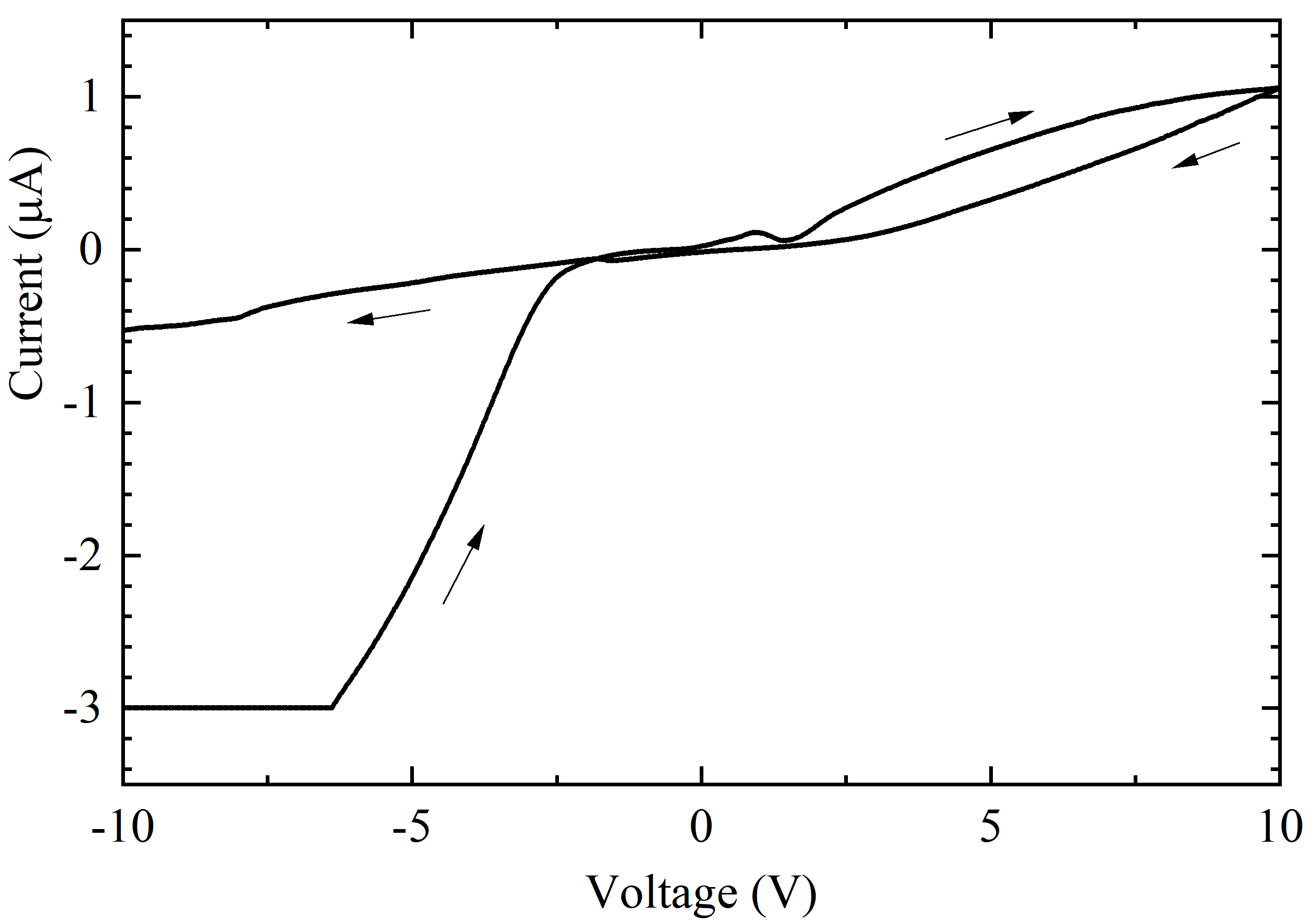}
    \caption{An IV trace showing the failure to arrive at an enduring BDT\textsuperscript{H2O} junction.}
    \label{fig:5}
\end{figure}

In the cases where the IV curve does not end up as shown in Fig.~\ref{fig:3}b or Fig.~\ref{fig:4}c the conductance may show a substantial initial increase in the IV curve. However at some point the conductance collapses to sub \si{\nano\ampere} values, as in the completely dry phase. This is shown in Fig.~\ref{fig:5}. The current compliance has been set at \SI{+3}{\micro\ampere} and \SI{-3}{\micro\ampere} for Fig.~\ref{fig:3}-\ref{fig:5}.

The possibility to compare same barrier different partially wet phase junctions or different barrier same partially wet phase junctions may provide an additional angle to increase our understanding of the physics of these systems. In comparing the IV curves from the BDT\textsuperscript{H2O} junction against a BDT\textsuperscript{THF} junction it should be noted that the BDT\textsuperscript{H2O} junction obtains a fixed and reproducible enduring state/IV-curve. For the BDT\textsuperscript{THF} junction this is not the case at ambient conditions. For the THF\textsuperscript{THF} or BDT\textsuperscript{THF} junction it is simple, at some point the THF microscopic layer will be evaporated. At ambient conditions there will not be an enduring BDT\textsuperscript{THF} or THF\textsuperscript{THF} junction. For H\textsubscript{2}O this is different as it is the microscopic layer at ambient conditions that lines everything \cite{bonn2009source1}. That it is possible to create an enduring BDT\textsuperscript{H2O} junction is fascinating. It implies that von Hippel’s vision is actually starting to become a reality \mcite{vonhippel1956source3a,*vonhippel1959source3b}. We now have the tools to create reproducible enduring single molecular junctions lined with a microscopic partially wet phase layer. The partially wet phase layer is an integral part of the total mechanism ensuring the stability of the entire system.

\section{Handling of the chemicals}
The molecular deposition on the electrodes of a freshly broken MCB junction starts by preparing the approximately \SI{1}{\milli\Molar} BDT/THF solution. The BDT molecules (\SI{99}{\percent} pure) are commercially available. For the THF solvent anhydrous BHT stabilized THF of \SI{99.85}{\percent} purity with less than \SI{50}{ppm} of water is used. The BDT molecules are handled in a home build miniature glove-box continuously flowed with Argon gas. This to ensure that the remaining BDT in the vial, used in future experiments, stays stored under Argon as well as the BDT used for the experiment ends up in a vial under Ar atmosphere. One vial of \SI{1}{\milli\Molar} BDT/THF solution per MCB assembly is prepared under Ar, using standard cannulation techniques for air sensitive reagents. The anhydrous THF is handled with a syringe, making use of a septum sealed THF bottle and septum sealed vials. Only at injecting the solution into the cell, by a glass syringe with stainless needle the solution is exposed to air.

\section{General discussion}
Weather can have an influence on the measurements. A low humidity (\SI{40}{\percent}) leads to considerable shorter measurement times. At low temperatures water may condensate for example on the cold cell wall which will lead to a higher number of unsuccessful junctions with a too low current start value in solution. The majority of measurements in ref \cite{muller2021source2} have been performed at temperatures around \SI{20}{\celsius} and at humidity’s in the \SI{50}{\percent} and \SI{60}{\percent}.

The use of a $\phi_\text{in}^{2}$ cell provides a more robust option against weather influences as compared to the $\phi_\text{in}^{4.5}$ cell. Just prior to the partially wet phase when there is only a minute amount of fluid in the cell left the evaporation rate in both cell types is of a similar order of magnitude. This is related to a small amount of fluid trapped somewhere, the amount being similar for both cell types. THF vapor outweighs air, so the $\phi_\text{in}^{2}$ cell with a 5 times smaller surface as compared to the $\phi_\text{in}^{4.5}$ cell is more robust against diffusion processes required for air to get to the bottom of the cell. 

On rare occasions an air bubble remains at the bottom of the cell, below the filled THF. Draining the last \si{\milli\meter} is in this case impossible as the entire THF amount will be gone once the bubble bursts. There is not a lot that can be done in this case. The junction is still in the partially wet phase once the bubble bursts. If it concerns a BDT barrier junction there is still a possibility to create a BDT\textsuperscript{H2O} junction by immediately raising the humidity to \SI{95}{\percent} followed by continuous IV scanning as described above. 

Successive use of an assembly for two or more times is recommended although often times not successful. As long as the current at \SI{-5}{\volt} has not been reduced below approximately \SI{-5}{\nano\ampere} a re-fill and re-drain procedure can be tried. If the current has fallen towards sub \SI{1}{\nano\ampere} levels there is no use in retrying. 

\section{Conclusions}
A breakdown of fabrication and process details has been provided for the creation and operation of a single molecule junction in the partially wet phase. Fabrication details describe material choices in combination with the used solvent and molecule. Process details can be used to arrive at the partially wet phase in a structured manner.


\bibliographystyle{unsrt}
\bibliography{references}

\end{multicols}

\end{document}